\documentstyle[eqsecnum,aps]{revtex}

\begin{document}

\title{Cosmological models with dynamical $\Lambda$ in scalar-tensor theories.}

\author{
L. M. D{\'\i}az-Rivera and L. O. Pimentel}

\address{ Department of Physics, \\ 
Universidad Aut\'onoma Metropolitana-Iztapalapa \\
Apdo. Postal 55-534, 09340 M\'exico, D.F, M\'exico }

\date{\today}

\maketitle
 
\begin{abstract}
In the context of a family of scalar-tensor theories with a dynamical $\Lambda$, that is a binomial on the scalar field,
the cosmological equations are considered. A general barotropic state equation $p=(\gamma-1)\rho$, for a perfect fluid is used for the matter content of the Universe. Some Friedmann-Robertson-Walker exact solutions are found, they have 
scale factor which shows exponential or power law dependence on time. For some models the singularity can be avoided. Cosmological parameters as $\Omega_m$, $\Omega_{\Lambda}$, $q_0$ and $t_0$ are 
obtained and compared with observational data.
\end{abstract}


\maketitle


\section{Introduction}
\label{Intro}

Unification theories have a nonzero cosmological constant that is about 120 orders of magnitude larger than the observed value for $\Lambda$, this constitutes the cosmological constant problem  \cite{weinberg}-\cite{Ng}.
In order to explain and solve such a problem, and to make compatible the actual observational 
data with the inflationary scenario and particle physics expectations, a time dependent 
cosmological constant was proposed \cite{bronstein}. This old idea has received a lot of 
attention (see e.g. \cite{oezer}-\cite{nesteruk}). What people have in mind is to make the 
vacuum energy dynamical. In such a way, during the evolution of the universe, the energy 
density of the vacuum decays into particles, thus leading to the decrease of the cosmological constant, obtaining as a result, although small, a creation of particles.  

A broad summary of cosmological models with time dependent cosmological ``constant'' is 
given by Overduin and Cooperstock \cite{overduin}, re-examining there, the evolution of 
the scale factor when $\lambda$ is given as function of $t, \, a(t), \, H,$ 
or $q$. A fairly general equation of state is considered and new numerical solutions are 
obtained, but as in most of the previous works, the time dependence of the cosmological 
term is introduced {\it ad hoc}. 

An alternative is an effective time dependent cosmological ``constant'' in the context of 
a scalar-tensor theories, which becomes a true constant for $t \gg 0$ \cite{capozziello97}.
Using Jordan-Brans-Dicke theory (JBD) in particular, the ``graceful exit'' problem of old 
inflationary cosmology might be improved. It remains the problem of determining  the 
JBD parameter $\omega$, that according to solar system experiments its value is 
$\|\omega\| \approx 500$, which has been derived from timing experiments using the Viking 
space probe \cite{reasenberg}. A better estimation of this parameter should be obtained 
from measure of others cosmological parameters in order to constrain $\omega$ more 
strongly than by means of solar system experiments \cite{barrow}. However, theories of very early 
universe as string theory, are better described in the context of JBD, and shows that 
$\omega$ can take negative values \cite{dahia}. 

Thus, scalar-tensor theories, and in particular JBD, are better theories, in order to 
get, in a natural way, a time dependent cosmological constant. Clearly, recent observational 
results restrict this kind of theory, e.g. the type Ia super nova (SN Ia) results, which in 
1998 show that $\Omega_{\Lambda} \sim 0.6$ \cite{perlmutter98} implying that our 
universe is speeding up. Thus, a model which attempts to describe the cosmological 
constant behavior, should take into account the observational evidences.

In a recent work \cite{coasting} we investigated the effect of a time dependent 
cosmological constant, in a family of scalar-tensor theories. There, we get cosmological 
models in the coasting period, where the time dependence on the cosmological constant 
occurs in a natural way. In  such a models we assumed a simple relation $\lambda(\phi)=
c \phi(t)^n$, (with $c$ and $n$ constants).

The existence of inflationary phase in scalar-tensor theories (STT) has been investigated 
by Pimentel and Stein-Schabes \cite{Pim-Stein}, finding  inflationary phases  for a 
polynomial cosmological constant in a general STT, which includes Brans-Dicke model with 
non-zero cosmological constant. On the other hand, Guendelman \cite{guendelman} 
has investigated the requirements of the potentials in order to have scale invariance. There was found 
the form of the potential needed by the global invariance, which in addition, its energy in the 
conformal Einstein frame has the characteristics for a suitable inflationary universe and  
$\Lambda$ decaying scenario for the late universe.

Motivated by these ideas, we shall consider a general STT as in our previous work 
\cite{coasting}, but now we shall consider a binomial $\lambda$ function on 
$\phi(t)$, in order to obtain exact solutions of the field equations, from which we 
obtain some kind of inflationary cosmological models and related cosmological 
parameters. In fact, we obtain in most of our solutions, a power law growth for the 
cosmological scale factor $a(t) \sim t^{\sigma}$, where $\sigma \gtrsim 1$ implies 
inflationary models. As it is known, this is a generic feature of a class of models that 
attempt dynamically to solve the cosmological constant problem. In our models $\sigma$ is 
a free parameter (at least in most of our models), in order to be adjusted by 
physical conditions and to be in agreement with recent data for SN Ia, that implies 
$\sigma \approx 1$, and which is consistent with the nucleosynthesis \cite{sethi99}.

Most of our solutions predict an accelerated expansion, such solutions are in agreement 
with the SN Ia results, but $\Omega_m$ and $\Omega_{\Lambda}$, depend on free parameters of 
our model. In some specific cases we get solutions with exponential growth of the scale 
factor.

In section \ref{fieldeq} we obtain the field equations and introduce our main ansatz. 
Section \ref{density} is devoted to obtain an expression of 
the density parameter and the corresponding contributions of the matter and scalar field. 
In section \ref{vacuum} we consider the vacuum case and obtain a set of exact solutions as 
well as cosmological parameters. In section \ref{baro} we consider the general case of a 
barotropic equation of state and obtain exact solutions for this general case. 
As example, we calculate specifically the case of a dust fluid and a stiff matter fluid. 
Also, we get and discuss the solutions of the radiation case in section 
\ref{radiation} and false vacuum case in section \ref{false-vacuum}. Finally in section 
\ref{conclusions} we summarize our results.

\section{ Field Equations}
\label{fieldeq}
We start with the action for the most general scalar-tensor theory of gravitation 
\cite{will}

\begin{equation}
\label{action}
S={1\over{16\pi G}}\int{d^4x\sqrt{-g}\,[\phi R-\phi^{-1}\omega g^{\mu\nu}
\partial_{\mu}\phi\partial_{\nu}\phi +2 \phi \lambda (\phi )]}+S_{NG},
\end{equation}

\noindent where $g=$ det ( $g_{\mu\nu} $), $G$ is Newton's constant,   
$S_{NG}\quad$ is the action for the non-gravitational matter. We   
use the signature  $(-,+,+,+)$. The arbitrary functions $\omega (\phi )$ and
$\lambda (\phi )$ distinguish the different scalar-tensor theories of
 gravitation; $\lambda (\phi )$ is a potential function and plays the role of a  
cosmological constant, $\omega(\phi) $ is the coupling function of the particular 
theory. 

The explicit field equations are

\begin{equation}
\label{field1}
G_{\mu \nu}={8 \pi T_{\mu \nu} \over {\phi}} + \lambda (\phi )\Box  
g_{\mu \nu} + \omega \phi^{-2}(\phi_{,\mu}\phi_{,\nu} - {1\over  
2}g_{\mu\nu}\phi_{,\lambda}\phi^{,\lambda})+\phi^{-1}(\phi_{;\mu\nu}-
g_{\mu\nu}\Box\, \phi),
\end{equation}

\begin{equation}
\label{field2}
\Box \,   \phi+{1\over 2}\phi_{,\lambda}\phi^{,\lambda}{d\over d\phi }\ln
\Bigl( {\omega (\phi ) \over \phi}\Bigr)+{1\over 2}{\phi \over \omega (\phi )}
\Bigl[R+2{d\over d\phi }(\phi \lambda (\phi )) \Bigr]=0,
\end{equation}

\noindent where $G_{\mu \nu}$ is the Einstein tensor. The last  
equation can be substituted by

\begin{equation}
\label{field3}
\Box \,    \phi +{2\phi ^2 d\lambda /d \phi -2\phi \lambda (\phi ) \over 3 + 2
\omega (\phi )}={1\over 3+2\omega (\phi )}\left (\,8\pi T- {d\omega \over  
d\phi} \phi_{,\mu} \phi ^{,\mu} \right ), 
\end{equation}

\noindent where $T=T_{\mu}^{\mu}$ is the trace of the stress-energy tensor. In a 
previous work \cite{coasting} was demonstrated that the divergence-less condition of 
the stress-energy matter tensor is satisfied if the field equation (\ref{field2}) is
satisfied too, although our field equations are given by Eqs. (\ref{field1}) and 
(\ref{field3}).

In what follows we shall assume $\omega (\phi) = constant$, $\lambda=\lambda (\phi)$. 
The corresponding field equations with a perfect fluid for the matter content in the 
isotropic and homogeneous line element 

\begin{equation}
\label{rw}
ds^2=-dt^2+a^2(t)\left [ {dr^2\over 1-k r^2}+
r^2(d\theta^2+\sin^2\theta \, d\phi^2) \right ],
\end{equation}

\noindent will be considered. Thus the field equations are

\begin{equation}
\label{field11}
 3\Big({\dot a \over a}\Big)^2+{3k\over a^2}-\lambda(\phi) - \frac{8\pi \rho}{\phi}  
-{\omega \over 2}\Big({\dot \phi\over \phi}\Big)^2 +  3{\dot a\over a}{\dot \phi \over \phi}=0,
\end{equation}

\begin{equation}
\label{field12}  
-2{{\ddot a}\over a}-\Big({\dot a\over a}\Big)^2-{k\over a^2} + \lambda(\phi) - 
\frac{8\pi p}{\phi} - 
{\omega \over 2}\Big({\dot\phi\over \phi}\Big)^2 - {{\ddot \phi}\over \phi} - 
2{\dot a \over a}{\dot \phi\over \phi}=0,
\end{equation}

\begin{equation}
\label{field13}
 \Bigg[{{\ddot \phi}\over \phi}+3{\dot a\over a}{\dot\phi\over \phi}\Bigg](3+2\omega) - 
2\Big({\lambda-\phi {d\lambda \over d\phi}}\Big) - {8\pi \over \phi}(\rho-3 p)=0.
\end{equation}

\noindent Where we have assumed $\phi=\phi(t)$, and the derivatives respect $t$ are 
denoted by a dot.

Assuming a barotropic equation of state $p=(\gamma-1)\rho$ and transforming
to the time $\tau$ defined as

\begin{equation}
\label{transt}
\tau=\int{ \phi^{1/2} dt},
\end{equation}

\noindent the set of equations (\ref{field11})-(\ref{field13}) are rewritten in 
the following way

\begin{equation}
\label{field21}
3 \Big({a' \over a}\Big)^2+{3k \over a^2\phi}-{\lambda(\phi) \over \phi}-
{8\pi\rho \over \phi^2}-{\omega \over 2}\Big({\phi'\over \phi}\Big)^2+
3{a' \over a}{\phi'\over \phi}=0,
\end{equation}

\begin{equation}
\label{field22}
-2{a''\over a}-{\phi''\over \phi}-\Big({a'\over a}\Big)^2-
{1\over 2}(1+\omega)\Big({\phi'\over \phi}\Big)^2-3{a'\over a}{\phi'\over \phi}-
{k \over a^2\phi}+{\lambda(\phi)\over \phi}-{8\pi\rho(\gamma-1)\over \phi^2} =0,
\end{equation}

\begin{equation}
\label{field23}
(3+2\omega)\Big[ {\phi''\over \phi}+{1\over 2}\Big({\phi'\over \phi}\Big)^2+
3{a'\over a}{\phi'\over \phi}\Big]-
2\Bigg({\lambda(\phi)\over \phi}-{d\lambda(\phi)\over d\phi}\Bigg)-
{8\pi\rho(4-3\gamma)\over \phi^2}=0,
\end{equation}

\noindent where the derivatives respect to $\tau$ are denoted by a prime.

\noindent In what follows we shall consider two important assumptions:

\begin{equation}
\label{a-phi}
a \phi^m=\alpha,
\end{equation}

\begin{equation}
\label{lambda-phi}
\lambda(\phi)=\lambda_{1}\phi^{n_{1}}+\lambda_{2}\phi^{n_{2}},
\end{equation}

\noindent where $m$, $\alpha$, $\lambda_{1}$, $\lambda_{2}$, $n_{1}$ and $n_{2}$ 
are constants. The first assumption is a very well known one (see e.g. \cite{johri} and 
reference therein), and it has been used as a condition for the deceleration parameter to be 
constant for flat models in Brans-Dicke theory. Furthermore, with this condition our 
field equations simplify notoriously and allows us to obtain exact solutions. The second 
condition is the main assumption of the present work which is motivated by the cosmological 
no-hair theorem for scalar-tensor theories \cite{Pim-Stein}, in order to study inflationary 
solutions in a theory of gravitation with a naturally dynamical cosmological constant. In this 
work we always work in the Jordan frame, where $G$ is variable, however we could make a 
conformal transformation to the Einstein frame where $G$ is constant and we have General Relativity 
plus a minimally coupled scalar field, then our potential becomes an exponential one, i.e., 
$V_1+ V_2$ $ \sim \exp(\epsilon n_1 \phi_c)+ \exp(\epsilon  n_2 \phi_c)$ (where $\epsilon$ is a 
constant and $\phi_c$ is a canonically defined scalar field). This is the type of potential according 
to Guendelman \cite{guendelman}, that is necesary to have scale invariance in a theory of gravitation 
free of the cosmological constant problem, that is, one with an early expanding phase and a 
$\Lambda$-decaying for late times. Details of the conformal transformation for STT can be seen in 
Ref. \cite{Pim-Stein}. With these assumptions, from equations (\ref{field21})-(\ref{field23}) we get

\begin{equation}
\label{field31}
\Big({\phi'\over \phi}\Big)^2\Big[ 3m^2-3m-{\omega \over 2}\Big] + 
{3k\over \alpha^2}\phi^{2m-1}-\lambda_{1}\phi^{n_{1}-1}-\lambda_{2}\phi^{n_{2}-1}
-{8\pi \rho\over \phi^2} =0,
\end{equation}

\begin{eqnarray}
\label{field32}
{\phi''\over \phi}(2m-1)+\Big({\phi'\over \phi}\Big)^2\Bigg[-3m^2+m-{\omega\over 2}-
{1\over 2}\Bigg] -{k\over \alpha^2}\phi^{2m-1}+\lambda_{1}\phi^{n_{1}-1}+ \nonumber \\
\lambda_{2}\phi^{n_{2}-1}-{8\pi\rho(\gamma-1)\over \phi^2}=0,
\end{eqnarray}

\begin{eqnarray}
\label{field33}
{\phi''\over \phi}(3+2\omega)+\Big({\phi'\over \phi}\Big)^2(3+2\omega)\Big({1
\over 2}-3m\Big)-2\lambda_{1}\phi^{n_{1}-1}-2\lambda_{2}\phi^{n_{2}-1}+ \nonumber \\
2\lambda_{1}n_{1}\phi^{n_{1}-1}+2\lambda_{2}n_{2}\phi^{n_{2}-1}-
{8\pi\rho(4-3\gamma)\over \phi^2} =0.
\end{eqnarray}

In the following sections we shall find exact solutions on different cases, 
as well as cosmological parameters which allow us to compare with actual observations 
of today value of Hubble parameter $H_{0}$, the actual value of deceleration parameter 
$q_{0}$, the density parameter $\Omega_{m}$ as well as the value of the vacuum energy 
density parameter $\Omega_{\Lambda}$.

\section{The density parameter}
\label{density}
Before we compute exact solutions of the set of field equations (\ref{field31})- 
(\ref{field33}), we shall get a general equation for $\Omega_{m}$ and $\Omega_{\phi}$, 
according to our proposed model.

\noindent Assuming $k=0$, Eq.(\ref{field11}) is written as

\begin{equation}
\label{omega1}
1={8\pi G \over 3H^2} \Bigg[ {\lambda(\phi) \over 8\pi G} + {\rho \over G\phi} 
+ {\omega \over 16 \pi G} \Big( {\dot{\phi} \over \phi} \Big)^2 - 
{3H \over 8 \pi G}{\dot{\phi} \over \phi} \Bigg].
\end{equation}

\noindent Defining 

\begin{eqnarray}
\label{def-gamas}
\Omega_{m} &=& {8 \pi \rho_m \over 3H^2} \, {1 \over \phi}, \nonumber \\
\Omega_{\phi} &=& {1 \over 3H^2} \Big[ \lambda(\phi) + 
{\omega \over 2} \Big( {\dot{\phi} \over \phi} \Big)^2 - 3H{\dot{\phi} \over \phi} 
\Big],
\end{eqnarray}

\noindent then we get

\begin{equation}
\label{omega-rho}
1=\Omega_{m} + \Omega_{\phi}.
\end{equation}

\noindent Taking into account the proposed relation (\ref{a-phi}), we get 

\begin{eqnarray}
\label{def-newgamas}
\Omega_{m} &=& {8 \pi \rho_{m} \over 3m^2} {1 \over \phi} \Big( {\phi \over  \dot{\phi}} 
\Big)^{2},  \nonumber \\
\Omega_{\phi} &=& {1 \over 3m^2} \Big( {\phi \over \dot{\phi}} \Big)^{2} \lambda(\phi) + 
{\omega \over 6m^2} + {1 \over m}.
\end{eqnarray}

\noindent According to the SN Ia observations \cite{perlmutter98}, the favored value 
of $\Omega_m \sim 0.4 \pm 0.1$ is given as a constrain to a cosmological constant

\begin{equation}
\Omega_{\Lambda}={4 \over 3}\Omega_m + {1 \over 3} \pm {1 \over 6},
\end{equation}

\noindent this implies $\Omega_{\Lambda} \sim 0.85 \pm 0.2$. It means that the SN Ia results 
are sensitive to the acceleration of the expansion, and constrain $4\Omega_m /3 -
\Omega_{\Lambda}$, which corresponds to the acceleration parameter at the median redshift 
of this objects, $z \sim 0.4$. Then the combination $\Omega_0=\Omega_m + 
\Omega_{\Lambda}$ is constrained by the microwave background radiation (CBR) anisotropy.
So that, $\Omega_0 \sim 1 \pm 0.2$ obtained from COBE and other measurements 
(see e.g., \cite{lineweaver}), and together with $\Omega_m \sim 0.4$, define a concordance region 
for $\Omega_{\Lambda} \sim 0.6$, becoming the best fit for the universe model \cite{turner91}, \cite{turner99}. 

In what follows, we shall compute both parameters of density in addition to the exact solutions of 
our field equations, in the different cases which we consider in this work.
\section{The vacuum case}
\label{vacuum}
Considering a vacuum case ($\rho=0$), we get from Eqs. (\ref{field31})-(\ref{field33}) 
the corresponding set of equations

\begin{equation}
\label{fvac1}
\Big({\phi'\over \phi}\Big)^2\Big[ 3m^2-3m-{\omega \over 2}\Big] + 
{3k\over \alpha^2}\phi^{2m-1}-\lambda_{1}\phi^{n_{1}-1}-\lambda_{2}\phi^{n_{2}-1} =0,
\end{equation}

\begin{equation}
\label{fvac2}
{\phi''\over \phi}(2m-1)+\Big({\phi'\over \phi}\Big)^2\Bigg[-3m^2+m-{\omega\over 2}-
{1\over 2}\Bigg] -{k\over \alpha^2}\phi^{2m-1}+
\lambda_{1}\phi^{n_{1}-1}+ \lambda_{2}\phi^{n_{2}-1}=0,
\end{equation}

\begin{eqnarray}
\label{fvac3}
{\phi''\over \phi}(3+2\omega)+\Big({\phi'\over \phi}\Big)^2(3+2\omega)\Big({1
\over 2}-3m\Big)-2\lambda_{1}\phi^{n_{1}-1}-2\lambda_{2}\phi^{n_{2}-1}+
2\lambda_{1}n_{1}\phi^{n_{1}-1}+ \nonumber \\
2\lambda_{2}n_{2}\phi^{n_{2}-1} =0.
\end{eqnarray}

\noindent Naturally for vacuum, the energy density is due to the contribution of the 
scalar field $\phi(t)$: $\Omega_{\phi}=1$. We found exact solutions of this set of 
equations in the following cases:

\begin{enumerate}
\item $k=0$
\begin{itemize}
\item $m=1/2$ This solution is not relevant for our purpose because $\lambda(\phi)$ becomes 
null.

\item $m=2/3$

\begin{eqnarray}
\label{svac-k0-m2/3-t}
\phi(t) &=& \phi_{1} \, t^{1/2} \nonumber \\
a(t) &=& a_{1}t^{-1/3} \nonumber \\
\lambda(\phi) &=& \lambda_{1} \, \phi(t)^{-4}, \nonumber \\
\omega &=& -7/3,    
\end{eqnarray}

\noindent where $\phi_{1}=\sqrt{2c_{1}}$, $a_{1}=\alpha \, (2c_{1})^{-1/3}$, 
$\lambda_{1}={c_{1}^2 / 2}$, $ \lambda_{2}=0$, and $c_{1}$ is an integration 
constant. This solution was written directly in terms of time $t$, according to 
(\ref{transt}). Here the expansion factor is decaying with the time, in conflict with observations, and $\omega$ has a negative values. A discussion of the meaning 
of negative values of $\omega$ is given in \cite{dahia}.

The Ricci scalar for this case is given by the following expression

\begin{equation}
\Re = {10 \over 3} \, t^{-2},
\end{equation}

\noindent where we can see that there is an initial singularity. 
The today deceleration parameter has a negative value, i.e., the present solution is an 
accelerated cosmological model

\begin{equation}
q_{0}=-4, \, \, \, \, \, \, \, \,   H_{0}=-{1 \over 3} \, t_{0}^{-1}, 
\end{equation}

\noindent according to this solution, the actual Hubble parameter has negative values, then we conclude that this model has not physical meaning.

\item $m \neq 1/2,\,\, 2/3$

On this case we get two families of solutions:

\begin{enumerate}
\item $\omega \neq { {3m^2+m-2} \over {2-3m} }$. 

\begin{eqnarray}
\label{svact-k0-mn1/2}
\phi(t)&=& \phi_{1} \, t^{2 \sigma}, \nonumber \\
a(t)&=& a_{1} \, t^{-2m \sigma}, \nonumber \\
\lambda(\phi)&=& \lambda_{1} \, \phi(t)^{-{1 \over \sigma}}, 
\end{eqnarray}

\noindent where 

\begin{eqnarray}
\mu &\neq & -2\nu, \nonumber \\
\sigma &=& \mu /(\mu + 2 \nu), \nonumber \\
a_{1} &=& \alpha \phi_{1}^{-m}, \nonumber \\ 
\phi_{1} &=& \big[c'^{1+{\nu \over \mu}}(\mu + 2 \nu)/2\big]^{{2\mu \over {\mu + 2\nu}}}, \nonumber \\
\lambda_{1} &=& (3m^2-3m-\omega/2)\mu^2 c'^{2(1+{\nu \over \mu})}, \, \, \, \, \,  
\, \, \lambda_{2 }= 0, \nonumber \\
\mu &\equiv & 12m^2-14m+4, \nonumber \\ 
\nu & \equiv &  -12m^2+5m+4\omega-6m\omega+2, 
\end{eqnarray}

\noindent and $c'$ is an integration constant. According to the solution of the 
present case, $\lambda$ is a monomial function on $\phi(t)$, although a time decaying one. 
Clearly the present solution requires $m \sigma < 0$ in order to be expanding. 
 
On the other hand, the Ricci scalar is given by

\begin{equation}
\label{revac-k0-mn1/2}
\Re= 12 m \sigma (4m\sigma +1) t^{-2}.
\end{equation}

\noindent According to this expression, the corresponding solution for the curvature scalar has an initial singularity. For this model, the present deceleration and Hubble 
parameter, are

\begin{equation}
\label{qvac-k0-mn1/2}
q_{0}=-1-{1 \over 2m \sigma}, \, \, \, \, \, \, \,   
H_{0}=-2m \sigma \, (t_{0})^{-1}, \, \, \, \, \, \, \,
t_0= {1 \over H_0} 2m \|\sigma\|.
\end{equation}

From these last equations we can see that the present model expands with acceleration if 
$m \sigma < -1/2 $. On the other hand, assuming $H_0 \sim 65 \pm 5$ km ${\rm s}^{-1}{\rm 
Mpc}^{-1}$ \cite{chaboyer98}, we get $t_0 \sim  15.05 \pm 1.96 \, \, \|m\sigma \|$ Gy, then 
the estimated age from this model is small compared with actual accepted values of 
$t_0$.

\item $\mu = -2 \nu$ $\Longrightarrow$ $\omega = { {3m^2+m-2} \over {2-3m} }$. 

For this particular relation between $\omega$ and $m$, we get the following solution

\begin{eqnarray}
\label{svact-k0-mn1/2-w}
\phi(t) &=& c' \, {\rm exp}[\phi_{2} t], \nonumber \\
a(t) &=& a_{2} \, {\rm exp} [-m \phi_{2} t], \nonumber \\
\lambda(\phi) &=& \lambda_{1} \phi(t),
\end{eqnarray}

\noindent where $\phi_2= c'^{1/2} \mu $, $a_{2}=\alpha c'^{-m}$, 
$\lambda_{2}=0$, $\lambda_{1}=4c'(3m^2-3m-\omega/2) \nu^2$ and $c'$ is an integration constant. In this case we 
get an inflationary exponential solution provided that $m \phi_2 < 0$ $\Longrightarrow$ 
$m < 0$ or $1/2 < m < 2/3$. This model is non-singular, as we can see from the Ricci scalar

\begin{equation}
\Re = 24 c' \, m^2 (6m^2-7m+2). 
\end{equation}

\noindent The present deceleration and Hubble parameters are given by

\begin{equation}
q_{0}= -1, \, \, \, \, \, \, \, \, \, \,  H_{0}= c'\, m(12m^2-14m+4),
\end{equation}

\noindent thus, this model expands with constant acceleration from a non-singular 
state, and with constant Hubble parameter. 

\end{enumerate}
\end{itemize}

\item $k \neq 0$

\begin{itemize}

\item $m=1/2$

\begin{eqnarray}
\label{svact-kn0-m1/2}
\phi(t)&=& \phi_{1} \, t^{-2}, \nonumber \\
a(t)&=& a_{1} \, t, \nonumber \\
\lambda(\phi) &=& \lambda_{1} \phi(t), 
\end{eqnarray}

\noindent where

\begin{eqnarray*}
\phi_{1} &=& \alpha^2{{3+2\omega} \over k}, \, \, \, \, \, \, \, \, \, \, \, \, \, \,  a_{1}=\Big({k \over 
{3+2\omega}}\Big)^{1/2},  \\
\lambda_{1} &=& {2k \over \alpha^2}, \, \, \, \, \, \, \, \, \, \, \, \, \, \, \, \, \, 
\, \, \, \, \, \, \, \, \, \, \, \, \,  \lambda_{2}=0. 
\end{eqnarray*}

\noindent According to this solution, $a(t)$ grows linearly with the time at a constant 
rate. The cosmological ``constant'' $\lambda$ decreases with the time. In order to have a 
real $a(t)$, we must have $k/(3+2\omega)>0$. This is a coasting cosmological 
solution which has initial singularities as it is shown from the corresponding Ricci 
scalar

\begin{equation}
\label{revac-kn0-m1/2}
\Re=12(3+2\omega) \, t^{-2}.
\end{equation}

\noindent On the other hand, as we have said, the today deceleration parameter becomes 
null, and the Hubble parameter is given by

\begin{equation}
\label{qvac-kn0-m1/2}
q_{0}=0, \, \, \, \, \, \, \, \, \, H=t_{0}^{-1},
\end{equation}

\noindent then, $t_0 = 1/H_0 \sim 15.05$ Gy, in relative agreement with actual 
observations.

\end{itemize}
\end{enumerate}
\section{Exact solutions for the case with a barotropic equation of state}
\label{baro}
In what follows we shall consider $\rho \neq 0$, so that returning to the set of 
equations (\ref{field31})-(\ref{field33}), we get from Eq. (\ref{field31})

\begin{equation}
\label{baro-rho}
{8\pi\rho\over \phi^2}=\Big({\phi'\over \phi}\Big)^2\Big[3m^2-3m-
{\omega\over 2}\Big]+{3k\over \alpha^2}\phi^{2m-1}-
\lambda_{1}\phi^{n_{1}-1}-\lambda_{2}\phi^{n_{2}-1}.
\end{equation}

\noindent Using this expression in equations (\ref{field32}) 
and (\ref{field33}), the set of field equations are reduced to the following two 
field equations

\begin{eqnarray}
\label{barof1}
&& {\phi''\over \phi}(2m-1)+\Big({\phi'\over \phi}\Big)^2\Bigg[-2m-\omega-
{1\over 2}-\gamma\Big(3m^2-3m-{\omega\over 2}\Big)\Bigg] + \nonumber \\
&& {(2-3\gamma)k\over \alpha^2}\phi^{2m-1}+
\gamma(\lambda_{1}\phi^{n_{1}-1}+\lambda_{2}\phi^{n_2-1})=0
\end{eqnarray}

\begin{eqnarray}
\label{barof2}
{\phi''\over \phi}(3+2\omega)+\Big({\phi'\over \phi}\Big)^2\Bigg[-12m^2+3m-
6\omega m+3\omega+{3 \over 2}+3\gamma\Big(3m^2-3m-{\omega\over 2}\Big)\Bigg] \nonumber \\
+ 3(3\gamma-4){k\over \alpha^2}\phi^{2m-1}+
(2-3\gamma)(\lambda_{1}\phi^{n_1-1}+\lambda_{2}\phi^{n_2-1})+
2(\lambda_1n_1\phi^{n_1-1}+\lambda_2\phi^{n_2-1})=0
\end{eqnarray}

\noindent From equation (\ref{barof1}) with $\gamma \neq  0$ (the false vacuum case
will be consider in section \ref{false-vacuum} ), we have

\begin{eqnarray}
\label{baro-lambda}
\lambda_1\phi^{n_1-1}+\lambda_2\phi^{n_2-1}=&&{{1-2m} \over \gamma}{\phi'' \over \phi}-
{1 \over \gamma}\Big[ -2m-\omega-{1\over 2}-\gamma \Big(3m^2-3m-
{\omega \over 2}\Big)\Big]{\phi'^2 \over \phi^2}- \nonumber \\
&&{(2-3\gamma)k \over \gamma \alpha^2}\phi^{2m-1}.
\end{eqnarray}

\noindent Using this expression and its derivative in Eq. (\ref{barof2}) we get the 
following equation

\begin{eqnarray}
\label{barof}
\lefteqn{ {2\over \gamma}(1-2m){\phi''' \over \phi}{\phi \over \phi'}+
{\phi'' \over \phi}\Big[{4 \over \gamma}(m+\omega+1)+6m(2m-1)\Big]+}\nonumber \\
&&{\Big({\phi' \over \phi}\Big)^2\Big[-12m^2-3m-6m\omega\Big]+ 
{k \over \alpha^2}\phi^{2m-1}\Big[12m-{4\over \gamma}(2m+1)\Big]}=0.
\end{eqnarray}

\noindent In order to solve this differential equation, we shall consider the 
value $m=1/2$, so that equation (\ref{barof}) is reduced to the following one 

\begin{equation}
\label{barofin}
{\phi'' \over \phi}-{3\gamma \over 4}\Big({\phi' \over \phi}\Big)^2+
{{3\gamma-4} \over {3+2\omega}}{k \over \alpha^2}=0.
\end{equation}

\noindent From this differential equation we have the two possible cases: $k=0$ and 
$k \neq 0$. The case $\gamma=4/3$ will be consider in section \ref{radiation}.

\begin{enumerate}

\item For $k=0$ we get the following solution which we write in terms of time $t$ as

\begin{eqnarray}
\label{solbaro-k0-t}
\phi(t)&=& \phi_{1} \, t^{\sigma}, \nonumber \\ 
a(t)&=& a_{1} \, t^{-\sigma/2},  \nonumber \\
\lambda(\phi)&=& \lambda_{1} \phi(t)^{-2/\sigma}, \nonumber \\
\rho &=& \rho_{1} a(t)^{-3 \gamma},
\end{eqnarray}

\noindent where 

\begin{eqnarray*}
\phi_1 &=& c_1 \Big[- {{3\gamma-2} \over {3\gamma-4}}c_1^{1/2} \Big]^{4 \over {2-3\gamma}}, 
\nonumber \\ 
\sigma &=& {4 \over {2-3\gamma}}, \\
a_{1} &=& {\alpha \over c_1^{1/2}}\Big[ -{{3\gamma-2} \over {3\gamma-4}} c_1^{1/2} \Big]^{
2 \over {3\gamma-2}}, \nonumber \\ 
\rho_1 &=& {1 \over \pi}{ {1-2\omega} \over \gamma (4-3\gamma)^2 }c_1^{2-{3\gamma \over 2}}, 
\alpha^{3\gamma} \nonumber \\
\lambda_{1} &=& \Big[ \omega \big( {1\over \gamma}-{1\over 2} \big)-{1\over 2}\big({1 \over 
\gamma}+{3 \over 2}\big)\Big] \Big({-{4 \over {4-3\gamma}}\Big)^2 c_1^{2-{3\gamma \over 2}},
\, \, \, \, \, \, \, \lambda_2=0}, 
\end{eqnarray*}

\noindent and $c_{1}$ is an integration constant. From this solution 
the expansion condition is $\sigma < 0$ $\Longrightarrow$ $\gamma > 2/3$.

\noindent In this case, the Ricci scalar and deceleration and 
Hubble parameters are given 
by the following expressions

\begin{equation}
\label{re-baro-k0}
\Re=36 {(2-\gamma) \over (3\gamma-2)^{2}}{1 \over t^{2}},
\end{equation}

\begin{equation}
\label{q0-baro-k0}
q_{0}={3 \over 2}\gamma-2 , \, \, \, \, \, \, \, \, \, \, 
H_{0}={2 \over {3\gamma-2}}{1 \over t_{0}}, \, \, \, \, \, \, \, \, \,  \Longrightarrow 
\, t_0= {2 \over {3\gamma-2}} {1 \over H_0}.
\end{equation}

\noindent Causality requires $0 \leq \gamma \leq 2$, so that this is an accelerated model 
for $\gamma < 4/3$, just at $\gamma=4/3$, $q_0=0$. On the other hand, $t_0 \sim 1/H_0 \sim 
15.05$ Gy for $\gamma \sim 4/3$. Then this is a kind of solution where the cosmic 
expansion is driven by the big-bang impulse.

\noindent The energy density parameter and the contribution of the scalar field $\phi(t)$ 
are given as follow

\begin{eqnarray}
\Omega_{m} &=& {2 \over 3\gamma} - {4\omega \over 3\gamma }, \nonumber \\
\Omega_{\phi} &=& {4\omega \over 3\gamma} - {2 \over 3\gamma} + 1.
\end{eqnarray}

\noindent In order to have a positive values of $\Omega_m$, $\omega < 1/2$ is required, 
including negative values of $\omega$. On the other hand, $\Omega_m \sim 0.4 \pm 0.1$, 
and $0 \leq \gamma \leq 2$, from our equations for $\Omega_m$ and $\Omega_{\phi}$, 
we get a restriction for $\omega$: $-0.1 \leq \omega \leq 1/2$. 

\item Considering now the case $k \neq 0$ from Eq. (\ref{barofin}), we get the following 
solution in terms of the parameter $\tau$:

\begin{itemize}

\item for ${k \over {3 + 2 \omega}}>0$

\begin{equation}
\label{sol-phi-kn0-tau-plus}
\phi(\tau)=c_{1}\Big[ \cosh \, \big( \beta \tau \big) \Big]^{\sigma}, \, \, \, \, 
k \neq 0,\, \gamma \neq 0, 4/3
\end{equation}

\noindent where now $\sigma={4 \over {4-3\gamma}}$, $\beta={(3\gamma-4) \over 
2\alpha}\sqrt{{k\over {3+2\omega}}}$, and $c_{1}$ is an integration constant. 
According to Eq. (\ref{a-phi}) we get

\begin{equation}
\label{sol-a-kn0-tau-plus}
a(\tau)= a_{1} \Big[ \cosh \, \big(\beta \tau \big) \Big]^{- {\sigma \over 2}}, \,
 \, \, \, k \neq 0,\, \gamma \neq 0, 4/3
\end{equation}

\noindent The Ricci scalar on this case is given by

\begin{equation}
\label{re-kn0-tau-plus}
\Re =r_1(r_2 + r_3)\cosh^{\sigma} \, [\beta \tau] - 
r_1r_2 \cosh^{\sigma-2} \, [\beta \tau].
\end{equation}

\noindent where $a_{1}=\alpha c_{1}^{-1/2}$, $r_{1}={3c_{1} \over \alpha^2}
{k \over {3+2\omega}}$, $r_{2}=3(2+\gamma)$, and $r_{3}=3\gamma + 2 + 4\omega$.
In order to know the singularities of this solution, we calculate the nonzero 
curvature invariant \cite{carminati}, which for this case, are given by

\begin{eqnarray}
\label{invars}
R_{1}&=&{3 \over 4}\Big[ {\ddot{a} \over a}-\Big( {\dot{a} \over a} \Big)^2-
{k \over a^2} \Big]^2, \nonumber \\
R_{2}&=&-{1 \over \sqrt{3}}R_{1}^{3/2}, \nonumber \\
R_{3}&=& {7 \over 12} R_{1}^{2},
\end{eqnarray}

\noindent then, it is enough to calculate $R_{1}$:

\begin{eqnarray}
\label{r1-invars-p} 
R_{1}&=&s_1(s_2-s_3)^2 \, \cosh^{2\sigma-4} \, [\beta \tau] +
s_1(s_3 + s_4)^2 \, \cosh^{2 \sigma} \, [\beta \tau] + \nonumber \\
&& 2s_1(s_2-s_3)(s_3 + s_4) \cosh^{2\sigma - 2} \, [\beta \tau].
\end{eqnarray}

\noindent where 

\begin{eqnarray*}
s_{1} &=& {3 \over 4}{c_{1}^{2} \over {\alpha^4 (3 + 2\omega )^2}}, \\
s_{2} &=& k \big(2-{3\over 2}\gamma \big), \\ 
s_{3} &=& k, \\ 
s_{4} &=& k (3 + 2\omega).
\end{eqnarray*}
 
\noindent From equation (\ref{re-kn0-tau-plus}), $\Re \rightarrow \infty $ provided that 
$\tau \rightarrow \pm \infty$, and at least one exponent is   positive, i.e.,
$\sigma > 0$ or $\sigma > 2$. On the other hand, from the curvature invariant, Eq.
(\ref{r1-invars-p}), $R_{1} \rightarrow \infty $ requires that $\tau \rightarrow \pm 
\infty$, and $\sigma >2$ ($\gamma > 2/3$), $\sigma > 0$($\gamma < 4/3)$ or $\sigma >1$ 
($\gamma > 0$). So that the solutions of this case, are singular for $0< \gamma< 4/3$, and 
$\tau \rightarrow \pm \infty$.

\item for ${k \over {3 + 2 \omega}}<0$

\begin{equation}
\label{sol-phi-kn0-tau-m}
\phi(\tau)=c_{1}\Big[\cos\, \big( \beta \tau \big) \Big]^{\sigma}, 
\, \, \, \, k \neq 0,\, \gamma \neq 0, 4/3
\end{equation}

\begin{equation}
\label{sol-a-kn0-tau-m}
a(\tau)=a_{1} \Big[ \cos \, \big( \beta \tau \big) \Big]^{-{\sigma \over 2}}, 
\, \, \, \, k \neq 0,\, \gamma \neq 0, 4/3
\end{equation}

\noindent where $\sigma$ and $a_{1}$ are defined as in the paragraph under 
equation (\ref{sol-phi-kn0-tau-plus}) and (\ref{re-kn0-tau-plus}). In this case,  
$\beta = {{3\gamma-4} \over 2\alpha}\sqrt{ \big\| {k \over {3+2\omega}} \big\|}$.

\begin{equation}
\label{re-kn0-tau-m}
\Re= r_1(r_3 - r_2) \cos^{\sigma} \, [\beta \tau] + 
r_1 r_2 \cos^{\sigma-2} \, [\beta \tau],
\end{equation}

\noindent where now 

\begin{eqnarray*}
r_{1} &=& {3c_{1} \over \alpha^2}\Big\|{k \over {3+2\omega}}\Big\|, \\ 
r_{2} &=& 3(2-\gamma), \\
r_{3} &=& (4-3\gamma)+2 \big\|3+2\omega\big\|.
\end{eqnarray*}

\noindent According to ({\ref{invars}}), as in the previous case, we need to calculate 
$R_{1}$ only, which for this case is given by

\begin{eqnarray}
\label{r1-invars-m} 
R_{1}&=&s_1(s_2 + s_3)^2 \, \cos^{2\sigma - 4} \, [\beta \tau] + 
s_1 (s_4 -s_3)^2 \cos^{2 \sigma} \, [\beta \tau] + \nonumber \\
&& 2s_1(s_2 + s_3)(s_4 - s_3) \cos^{2\sigma - 2} \, [\beta \tau],
\end{eqnarray}

\noindent where 

\begin{eqnarray*}
s_{1} &=& {3 \over 4}{c_{1}^{2} \over {\alpha^4|3 + 2\omega|^2}}, \\ 
s_{2} &=& \|k\|\big(2-{3\over 2}\gamma \big), \\ 
s_{3} &=& \|k\|, \\ 
s_{4} &=& k\|3+2\omega \|.
\end{eqnarray*}

\noindent From Eqs. (\ref{re-kn0-tau-m}) and (\ref{r1-invars-m}), $\Re \rightarrow 
\infty$ as well as $R_{1} \rightarrow \infty$, if $[\beta \tau]\rightarrow  
\pm (2n+1)\pi/2$ and at least one exponent on the respective expressions of 
$\Re$ and $R_1$ is negative: $\sigma < 2$ ($\gamma < 2/3$), $\sigma < 1$ ($\gamma < 0$) 
or $\sigma < 0$ ($\gamma > 4/3$). Then we have the two ranges of $\gamma$ for which 
the solutions of the present case, are singular: $0 < \gamma < 2/3$ and $4/3 < \gamma \leq 
2$, since furthermore causality requires $\gamma$ to be in the interval 
$0 \leq \gamma \leq 2$.
  
\end{itemize}

\noindent On the other hand, from equations (\ref{baro-rho}) and (\ref{baro-lambda}) we get 
respectively, for both possibilities ${k \over {3 + 2 \omega}}<0$ or 
${k \over {3 + 2 \omega}}>0$

\begin{equation}
\label{sol-rho-kn0-pm}
\rho=\rho_{1} a(\tau)^{-3\gamma},
\end{equation}

\begin{equation}
\label{sol-lambda-kn0-pm}
\lambda(\phi)=\lambda_{1} \phi(\tau)^{{3 \over 2}\gamma -1}+\lambda_{2}\phi(\tau),
\end{equation}

\noindent where 

\begin{eqnarray*}
\rho_1={ k c_{1}^{2-{3\gamma \over 2}} \over {4\pi \alpha^2 \gamma}} \alpha^{3\gamma}, \, \, \, \, \,   
\lambda_1={ k c_{1}^{2-{3\gamma \over 2}} \over \alpha^2 }\Big(1-{2 \over \gamma}\Big),
\, \, \, \, \, \lambda_{2}={2k \over \alpha^2}.
\end{eqnarray*}

\noindent As we can see, $\lambda(\phi)$ remains as a binomial 
function of $\phi$ if $\gamma \neq 4/3$. In order to analyze the behavior of the obtained 
solution in terms of the cosmological time $t$, we shall give some examples.

\end{enumerate}

\subsection{Dust fluid}
\label{dust}

\noindent One interesting application of a barotropic equation of state  corresponds to 
a dust fluid ($\gamma =1$), on that case the solution reads as follows
 
\begin{enumerate}

\item $k=0$

\begin{eqnarray}
\label{sol-dust-k0-t}
\phi(t)&=& \phi_{1} \, t^{-4}, \nonumber \\
a(t)&=& a_{1} \, t^{2}, \nonumber \\
\lambda(\phi)&=& \lambda_{1} \phi(t)^{1/2},  \nonumber \\
\rho&=& \rho_{1} a(t)^{-3}.
\end{eqnarray}

\noindent Where $\phi_{1}=c_{1}^{-1}$, $a_{1}=\alpha c_{1}^{1/2}$, 
$\lambda_{1}= 4(2\omega-5)c_{1}^{1/2}$, $\lambda_{2}=0$, 
$\rho_{1}={ {1-2\omega} \over \pi}c_{1}^{1/2}\alpha^3$, and $c_{1}$ is an integration constant. 
This is an extended inflationary  solution, with a time decaying cosmological constant 
and initial singularity, as it is shown by the corresponding Ricci scalar

\begin{equation}
\label{re-dust-k0-t}
\Re=36 \, t^{-2}.
\end{equation}

\noindent The expansion takes place with a constant acceleration

\begin{equation}
\label{qo-dust-k0}
q_{0}=-{1 \over 2}, \, \, \, \, \, \, \, \, \, \,  H_{0}={2 \over t_{0}}, \, \, \, \, \, \, \,  \Longrightarrow \, \, \,  t_0 \sim {2 \over H_0}.
\end{equation}

\noindent With $H_0 \sim 65 \pm 5$ km ${\rm s}^{-1}{\rm Mpc}^{-1}$, we obtain $t_0 \sim 30.1$ 
Gy, which is too big value compared with the globular cluster age.
 
\noindent The density parameters for a dust fluid are given by

\begin{eqnarray}
\Omega_{m} &=& {2 \over 3}- {4\omega \over 3} , \nonumber \\
\Omega_{\phi} &=& {4\omega \over 3} + {1 \over 3}.
\end{eqnarray}

\noindent On this case, as in the general case, in order to have a positive values of 
$\Omega_m$, it is required that $\omega < 1/2$, including negative values. 
On the other hand, according to observational results, $\Omega_m \sim 0.4 \pm 0.1$ and 
$\Omega_{\Lambda} \sim 0.6$, then $\omega$ is restricted to be $\omega \sim 1/5$.

\item $k \neq 0$

\begin{itemize}
\item For ${k \over {3+2\omega}} >0$

\noindent The solutions in terms of the time $t$ is given by 

\begin{eqnarray}
\label{sol-dust-kn0-plus-t}
\phi(t)&=&c_{1}[1-\phi_1 \, t^2]^{-2}, \nonumber \\
a(t)&=&a_{1}[1-\phi_{1} \, t^2] ,\nonumber \\
\lambda(\phi)&=& \lambda_{1} \phi(t)+ \lambda_{2} \phi(t)^{1/2} \nonumber \\
\rho&=& \rho_{1} a(t)^{-3}, 
\end{eqnarray}

\noindent where 

\begin{eqnarray*}
a_{1}=\alpha / c_{1}^{1/2}, \, \, \, \, \phi_{1}={c_1 \over 4 \alpha^2} 
{k \over {3+2\omega}}, \, \, \, \, \lambda_{1}={2k \over \alpha^2}, \, \, \, \, 
\lambda_{2}= -{kc_{1}^{1/2} \over \alpha^2}, \, \, \, \, \rho_{1}={k c_{1}^{1/2} 
\over {4\pi \alpha^2}} \alpha^{3}, 
\end{eqnarray*}

\noindent and $c_{1}$ is an integration constant. Clearly $a(t)$ 
and $\phi(t)$ must be positive, in order to be physically significant; this requirement restricts the range of values which $t$ can take: $t< \phi_1^{-1/2}$, and $c_1 > 0$ 
as we can see from the definition of $\phi_1$. In this case the cosmological term, in spite of being a binomial function on $\phi$, decays with the time.

The corresponding Ricci scalar and curvature invariant show that this solution is singular:

\begin{equation}
\label{r1-dust}
\Re={6 \over \Big[1 -\phi_{1} \, t^2 \Big]^2} 
\Bigg[6 \phi_{1}^2 \, t^2-2 \phi_{1}+{k \over a_{1}^{2}}\Big],
\end{equation}

\begin{equation}
\label{re-dust-kn0-plus-t}
R_{1}= {3 \over 4}{1 \over \Big[1 -\phi_{1} \, t^2 \Big]^4} 
\Bigg[-2 \phi_{1}^2 \, t^2-2 \phi_{1} - {k \over a_{1}^{2}}\Big]^2,
\end{equation}

\noindent According to our analysis of the general solution, for $\sigma=4/ 
(4-3\gamma)$, with $\gamma = 1$ we get $\sigma=4$. Taking into account the 
general equations of the Ricci scalar and curvature invariant, Eqs. 
(\ref{re-kn0-tau-plus}) and (\ref{r1-invars-p}), $\Re \longrightarrow \infty$ and 
$R_1 \longrightarrow \infty$ for $\sigma=4$ and $\tau \longrightarrow \pm \infty$, 
which corresponds to a finite value of $t$, according to the time dependent solution 
(\ref{sol-dust-kn0-plus-t}). Furthermore $\gamma=1<4/3$, is consistent with our 
singularity requirement in our discussion for the general case with 
${k \over {3+2\omega}}>0$.

\noindent The today values of the deceleration and Hubble parameters are given by the following expressions

\begin{equation}
\label{q0-dust-kn0-plus-t}
q_{0}={1 \over 2}\Big( {1 \over \phi_1 \, t_0^{2}}-1 \Big), \, \, \, \, \, \, \, \,
H_{0}={ 2 \phi_1 \, t_0 \over {\phi_1 \, t_0^2-1} }, \, \, \, \, \, 
t_0= {1 \over H_0} \pm \sqrt{ {1\over H_0^2}+{1 \over \phi_1} }.
\end{equation}

\noindent Because $\phi_1 t_0^2 < 1$, then $q_0 > 0$. This model expands from $t=-\phi_1^{-1/2}$ until $t=0$, then it contracts until $t=\phi_1^{-1/2}$, in both cases with positive deceleration parameter.
The numerical value for $t_0$ depends on the values of the free constants.

\item For ${k \over {3 + 2 \omega}}<0$

\noindent The corresponding solution in terms of the time $t$ is given as

\begin{eqnarray}
\label{sol-dust-kn0-t}
\phi(t)&=&c_{1}[1+\phi_{1} \, t^2]^{-2}, \nonumber \\
a(t)&=& a_{1}[1+\phi_{1} \, t^2], \nonumber \\
\lambda(\phi)&=& \lambda_{1} \phi(t)+\lambda_{2}\phi(t)^{1/2} \nonumber \\
\rho&=& \rho_{1} a(t)^{-3}, 
\end{eqnarray}

\noindent where 

\begin{eqnarray*}
\phi_{1}={c_{1} \over 4\alpha^2} \big\|{{k \over 3+2\omega}}\big\|, \, \, \, \, \, \,  
a_{1}={\alpha / c_{1}^{1/2}}, \, \, \, \, \, \, \lambda_{1}={2k \over \alpha^2}, \, \, \, \, \, \, \lambda_{2}=-{kc_{1}^{1/2} \over \alpha^2}, \, \, \, \, \, \, \rho_{1}={k c_{1}^{1/2} \over 
{4\pi \alpha^2}} \alpha^3,
\end{eqnarray*} 
  
\noindent and $c_{1}$ is an integration constant. In this case we have not 
restrictions on the values which $t$ can take. If $-1/\phi_1 < (t_0+c)^2$, the expansion takes place with non-constant acceleration and without singularity, as it is shown from the corresponding Ricci scalar and curvature invariant 

\begin{equation}
\label{re-dust-kn0-t-m}
\Re={6 \over \Big[1 +\phi_{1} \, t^2 \Big]^2} 
\Bigg[6 \phi_{1}^2 \, t^2+2 \phi_{1}+{k \over a_{1}^{2}}\Big],
\end{equation}

\begin{equation}
\label{r1-dust-kn0-plus-t}
R_{1}= {3 \over 4}{1 \over \Big[1 + \phi_{1} \, t^2 \Big]^4} 
\Bigg[-2 \phi_{1}^2 \, t^2 + 2 \phi_{1} - {k \over a_{1}^{2}}\Big]^2.
\end{equation}

\noindent According to our general analysis of this case, from (\ref{re-kn0-tau-m}) and 
(\ref{r1-invars-m}), $\Re$ and $R_1$ do not diverge for $\sigma=4$ and 
$[\beta \tau]\rightarrow \pm (2n+1)\pi/2$, in agreement with the time dependent 
solution (\ref{sol-dust-kn0-t}) which has not singularities. $\gamma=1 < 4/3$ in this  
model, which is consistent with our condition $\gamma > 4/3$ as requirement for existence of singularities.

For this solution, the corresponding present values of the density and Hubble parameters are given as

\begin{equation}
\label{q0-dust-kn0-t}
q_{0}=-{1\over 2}-{1 \over 2\phi_{1}} \, t_{0}^{-2}, \, \, \, \, \, \, \, \, 
H_{0}={ 2 \phi_{1} \, t_{0} \over {\phi_{1} \, t_{0}^2+1} }, \, \, \, \, \, 
t_0= {1 \over H_0} \pm \sqrt{ {1\over H_0^2}-{1 \over \phi_1} }.
\end{equation}

\noindent As in the previous case, the values of $t_0$ depends on the free constants 
of our model, but in this case, for $H_0^2 \sim \phi_1 \, \Longrightarrow t_0 \sim 1/H_0$. 

\end{itemize}
\end{enumerate}

\subsection{Stiff matter fluid}
\label{stiff}

Another interesting application of a barotropic equation of state is a stiff matter 
fluid for which $\gamma = 2$. In such a case, the solutions (\ref{solbaro-k0-t}),
(\ref{sol-phi-kn0-tau-plus}),({\ref{sol-a-kn0-tau-plus}), 
(\ref{sol-phi-kn0-tau-m}), and (\ref{sol-a-kn0-tau-m}) are the following ones

\begin{enumerate}

\item k=0

\noindent The solution in terms of the physical time $t$ is given by

\begin{eqnarray}
\label{sol-stiff-k0-t}
\phi(t)&=& \phi_{1}(c_-t)^{-1}, \nonumber \\
a(t)&=& a_{1}(c-t)^{1/2}, \nonumber \\
\lambda(\phi)&=& \lambda_{1} \phi(t)^{2},  \nonumber \\
\rho&=& \rho_{1} a(t)^{-6},
\end{eqnarray}

\noindent where now

\begin{eqnarray*}
\phi_{1}={c_{1}^{1/2} \over 2}, \, \, \, \, \,   
a_{1}={\alpha \sqrt{2} \over c_{1}^{1/4}}, \, \, \, \, \, \lambda_{1}=-{4 \over c_{1}}, \, \, \, \, \, \lambda_{2}=0, \, \, \, \, \, \rho_{1}={1 \over 8\pi}{{1-2\omega} \over c_{1}} \alpha^6.
\end{eqnarray*} 

\noindent This solution has a physical meaning  for $t < c$, where $c$ is an integration constant and the scale factor increases very slowly with the time, and with constant deceleration. The Ricci scalar and Hubble parameter, are given by

\begin{equation}
\label{re-stiff-k0-t}
\Re=0,
\end{equation}

\begin{equation}
\label{q0-stiff-k0-t}
q_{0}=1, \, \, \, \, \, \, \, \, \, \, \, H_{0}={1 \over 2}\, {1 \over {t_{0}-c}},
\end{equation}

\noindent Naturally this model is not valid today because it shrink for the allowed range of $t$.

\noindent The corresponding density parameters are

\begin{eqnarray}
\Omega_{m} &=& {1 \over 3}- {2 \omega \over 3}, \nonumber \\ 
\Omega_{\phi} &=& {2 \over 3} + {2\omega \over 3}.
\end{eqnarray}

\noindent In order to have a positive values of $\Omega_m$, then $\omega < 1/2$, as we have claimed in the discussion of the general solution. The observational results $\Omega_m = 0.4$ and $\Omega_{\Lambda}=0.6$, determine $\omega = -0.1$.

\item $k \neq 0$

\begin{itemize}

\item For ${k \over {3+2\omega}}>0$ 

\noindent The solution of this case, in terms of the time $t$ is given as

\begin{eqnarray}
\label{sol-stiff-kn0-plus-t}
\phi(t) &=& c_{1}\Big[ 1+ \phi_{1} \, t^2 \Big]^{-1}, \nonumber \\
a(t) &=& a_{1} \Big[ 1+ \phi_{1} \, t^2 \Big]^{1/2},   \nonumber \\
\lambda(\phi)&=& \lambda_{1} \phi(t),  \nonumber \\
\rho &=& \rho_{1} a(t)^{-6}, 
\end{eqnarray}

\noindent where $\phi_{1}={c_{1} \over \alpha^2}{k \over {3+2\omega}}$, ($\phi_1>0$ 
for $c_1 > 0$), $a_{1}=\alpha c_{1}^{-1/2}$, $\lambda_{1}={2k \over \alpha^2 }$, 
$\lambda_{2}=0$, $\rho_{1}={k \over 8\pi c_{1}} \alpha^4$, and $c_{1}$ is an 
integration constant. Here $a(t)$ grows with the time, from a minimum radius $a_1$ there is a non-singular state. The expansion takes place with non-constant acceleration, as we can see 
from the corresponding Ricci scalar and curvature invariant, which are given by  

\begin{equation}
\label{re-stiff-kn0-plus-t}
\Re={6  \over \big[ 1+ \phi_{1} \,t^{2}\big]}
\Bigg(\phi_{1} + {kc_{1} \over \alpha^2}  \Bigg),
\end{equation}

\begin{equation}
\label{r1-stiff-kn0-plus-t}
R_{1}={3 \over 4}{1  \over \big[ 1+ \phi_{1} \,t^{2}\big]^4}
\Bigg[ -\phi_{1} t^2 \, \bigg(\phi_{1}+{c_1 k \over \alpha^2}\bigg) + \phi_1 - 
{kc_{1} \over \alpha^2} \Bigg]^2.
\end{equation}

\noindent In this case $\gamma=2$ implies $\sigma = -2$. from Eqs. (\ref{re-kn0-tau-plus}) and (\ref{r1-invars-p}) with this value of  $\sigma$, neither $\Re$ nor $R_1$ diverge for $\tau \longrightarrow \pm \infty$ which is consistent with our requirement $\gamma > 4/3$ for the  avoidance of singularities. This analysis is in agreement with the inspection of Eqs. (\ref{re-stiff-kn0-plus-t}) and (\ref{r1-stiff-kn0-plus-t}).

\noindent The deceleration and Hubble parameters are

\begin{eqnarray}
\label{q0-stiff-kn0-plus-t}
q_{0} &=& -{1 \over \phi_{1}} \, t_{0}^{-2}, \nonumber \\
H_{0} &=& {c_{1} \over \alpha^2} {k \over {3 + 2\omega}}  \,
{{t_{0}} \over {1 + {c_{1} \over \alpha^2}}{k \over {3 + 2\omega}} t_{0}^2 }, \nonumber \\
t_0 &=& {1 \over 2H_0} \pm \sqrt{ {1 \over 4H_0^2} - {\alpha^2 \over c_1} { 3+ 2\omega 
\over k} }.
\end{eqnarray}

\item For ${k \over {3 + 2 \omega}}<0$

\begin{eqnarray}
\label{sol-stiff-kn0-m-t}
\phi(t) &=& c_{1}\Big[ 1-\phi_{1} \, t^2 \Big]^{-1}, \nonumber \\
a(t) &=& a_{1} \Big[ 1- \phi_{1} \, t^2 \Big]^{1/2},   \nonumber \\
\lambda(\phi)&=& \lambda_{1} \phi(t), \nonumber \\
\rho &=& \rho_{1} a(t)^{-6}, 
\end{eqnarray}

\noindent where $\phi_1={c \over \alpha^2}\big\|{k \over {3 + 2 \omega}}\big\|$. 
In order to have a physically solution, it is required that $\phi_1 < 0$ 
$\Longrightarrow$ $c < 0$. The corresponding Ricci scalar and curvature invariant in 
the present case, are given by

\begin{equation}
\label{re-stiff-kn0-m-t}
\Re={6 \over \Big[ 1- \phi_{1} \, t^{2}\Big]}
\Bigg( {k c_{1} \over \alpha^2}- \phi_{1} \Bigg),
\end{equation}

\begin{equation}
\label{r1-stiff-kn0-m-t}
R_{1}={3 \over 4}{1  \over \big[ 1 - \phi_{1} \,t^{2}\big]^4}
\Bigg[ -\phi_{1} t^2 \, \bigg(\phi_{1}-{c_1 k \over \alpha^2}\bigg) - \phi_1 - 
{kc_{1} \over \alpha^2} \Bigg]^2.
\end{equation}

\noindent In this case $\gamma=2$, then  $\sigma=-2$; such that in the general solution ( Eqs. (\ref{sol-phi-kn0-tau-m})- (\ref{r1-invars-m})), $\Re \longrightarrow \infty$ and 
$R_1 \longrightarrow \infty$, i.e., the solution for this case is singular
provided that $[\beta \tau]$ $\Longrightarrow$ $\pm (2n+1)\pi/2$, which 
corresponds to $t= \phi_1^{-1/2}$. According to our singularity discussion under equation
(\ref{r1-invars-m}), for $\gamma = 2$ the corresponding solution should being singular, 
as we can verify from inspection of Eq. (\ref{re-stiff-kn0-m-t}) and 
(\ref{r1-stiff-kn0-m-t}).

\noindent the present deceleration and Hubble parameters are

\begin{eqnarray}
\label{q0-stiff-kn0-m-t}
q_{0} &=& {1 \over \phi_{1}} \, t_{0}^{-2}, \nonumber \\
H_{0} &=& -{c_{1} \over \alpha^2}\Big\| {k \over {3 + 2\omega}} \Big\| \,
{t_{0} \over {1 - {c_{1} \over \alpha^2}} \big\| {k \over {3 + 2\omega}} \big\| t_{0}^2}, 
\nonumber \\
t_0 &=& {1 \over 2H_0} \pm \sqrt{ {1 \over 4H_0^2} + {\alpha^2 \over c_1} { 3+ 2\omega 
\over k} }.
\end{eqnarray}

\noindent As we can see, $\phi_1 < 0$ $\Longrightarrow$ $q_0 <0$, then this model is accelerated. The numerical values of $H_0$ and $t_0$ depend on the values of the free  constants.

\end{itemize}

\end{enumerate}

\section{The radiation case}
\label{radiation}

We shall consider the radiation case for which $\gamma =4/3$, so that returning to 
Eqs. (\ref{field31})-(\ref{field33}) and following a similar procedure as in section 
\ref{baro}, we get Eq. (\ref{barof}) with $\gamma =4/3$. In order to solve this 
differential equation, we shall assume $m=1/2$, then we have for this case

\begin{equation}
\label{field1-radiation}
{\phi'' \over \phi}-\Big({\phi' \over \phi}\Big)^2=0,
\end{equation}

\noindent for which $\omega \neq -3/2$. This differential equation has the 
following solution

\begin{equation}
\label{sol-phi-radiation}
\phi(\tau)=c_1 e^{c \tau},
\end{equation}

\noindent where $c$ and $c_{1}$ are integration constants. According to Equations 
(\ref{a-phi}), (\ref{baro-rho}) and (\ref{baro-lambda}) with $\gamma =4/3$ and $m=1/2$ 
we get respectively

\begin{eqnarray}
\label{sol-radiation-tau}
a(\tau)&=& a_{1} e^{-c \tau /2}, \nonumber \\
\lambda(\phi)&=& \lambda_{1}\phi(\tau), \, \, \, \lambda_{2}=0, \nonumber \\
\rho&=& \rho_{1} a(\tau)^{-4},
\end{eqnarray}

\noindent where $a_{1}=\alpha c_{1}^{-1/2}$, $\lambda_{1}=
\Big[ {c^2 \over 8}(3+2\omega)+{3 \over 2}{k \over \alpha^2} \Big]$ and 
$\rho_{1}={3 \over 16 \pi}\Big[ {k\over \alpha^2}-{\omega c^2 \over 2}-{3 c^2 \over 4} 
\Big] \alpha^4$. In terms of the time $t$, this solution is given as follows

\begin{eqnarray}
\label{sol-radiation-t}
\phi(t)&=& \phi_{1} \, t^{-2}, \nonumber \\
a(t)&=& a_{1} \, t, \nonumber \\
\lambda(\phi)&=& \lambda_{1}\phi(t),  \nonumber \\
\rho&=& \rho_{1} a(t)^{-4},
\end{eqnarray}

\noindent where $\phi_{1}={4 \over c^2}$, $a_{2}={\alpha c \over 2}$ and $\lambda_{2}=0$. 
This is a singular solution according to the Ricci scalar which for this case is given by

\begin{equation}
\label{re-radiation-t}
\Re= 6 \Big(1+{k \over a_{1}^2} \Big) \, t^{-2}.
\end{equation}

\noindent The deceleration parameter becomes null, instead the today Hubble parameter 
is given as

\begin{equation}
\label{q0-radiation-t}
q_{0}=0, \, \, \, \, \, \, \, \, H_{0}={1 \over t_{0}},
\end{equation}

\noindent and therefore $t_0=1/H_0 \sim 15.05$ Gy.

\noindent The case $k=0$ is not excluded from the solution (\ref{sol-radiation-t}). For $k=0$ 
we get the density parameter from the matter and the scalar field as 

\begin{eqnarray}
\Omega_{m} &=& - \, \omega - {3 \over 2}, \nonumber \\
\Omega_{\phi} &=& \, \, \, \, \omega + {5 \over 2}.
\end{eqnarray}

\noindent It is required that $\omega < -3/2$, in order to have $\Omega_m > 0$, and the 
observational accepted values of $\Omega_m$ and $\Omega_{\phi}$, determine 
$\omega = -1.9$.

\section{The false vacuum case}
\label{false-vacuum}

We analyze now the case $\gamma =0$. We shall follow a similar procedure as in 
section \ref{baro}, then we get from Eqs. (\ref{barof1}) and (\ref{barof2}) with 
$\gamma=0$, the following set of equations

\begin{equation}
\label{field11-false-vacuum}
{\phi''\over \phi}(2m-1)+\Big({\phi'\over \phi}\Big)^2\Bigg[-2m-\omega-
{1\over 2}\Bigg]
+{2 k\over \alpha^2}\phi^{2m-1}=0,
\end{equation}

\begin{eqnarray}
\label{field21-false-vacuum}
{\phi''\over \phi}(3+2\omega)+\Big({\phi'\over \phi}\Big)^2\Bigg[-12m^2+3m-
6\omega m+3\omega+{3 \over 2}\Bigg]+ {12 k\over \alpha^2}\phi^{2m-1}+
2{\lambda(\phi)\over \phi}+2{d\lambda(\phi)\over d\phi}=0.
\end{eqnarray}

\noindent We consider first the case $k=0$

\begin{itemize}

\item $k=0$

\noindent The set of equations (\ref{field11-false-vacuum})-(\ref{field21-false-vacuum}) 
with $k=0$ has the two possible set of solution depending on the relation between $m$ and 
$\omega$:

\begin{enumerate}

\item $\omega = -1-m$

\begin{eqnarray}
\phi(t) &=& c_1 {\rm exp}[\phi_1 \, t], \nonumber \\
a(t) &=& a_1 {\rm exp} [-m \phi_1 \, t], \nonumber \\
\lambda(\phi) &=& {\lambda_1 \over \phi(t)} + \lambda_2, \nonumber \\
\rho &=& -{c_1 \over 8 \pi},
\end{eqnarray}

\noindent where 

\begin{eqnarray*}
\phi_1=c_3^{1/2}(2m-1), \, \, \, \, \, \, a_1=\alpha c_1^{-m}, \, \, \, \, \, \, 
\lambda_1=c_1, \, \, \, \, \, \, \lambda_2={c_1 \over 2}(2m-1)^3(3m-1).
\end{eqnarray*}

\noindent This is an inflationary solution if $m \phi_1 < 0$. This condition means 
$0 < m < 1/2$, which implies a condition on the range of the values of $\omega$: 
$1/2 < \omega < 1$. In order to have physical solutions, $c_1 > 0$, which means $\rho < 0$.
Of course, these solutions have not singularities, as we can see from the Ricci scalar

\begin{equation}
\Re = 12 c_1 m^2 (2m-1)^2.
\end{equation}

\noindent The deceleration and Hubble parameters are given by

\begin{equation}
q_0=-1, \, \, \, \, \, \, \, \, \, H_0=m(1-2m)c_1^{1/2},
\end{equation}

\noindent thus the model is accelerated. On the other hand, the density parameter due 
to the matter and scalar field, are 

\begin{eqnarray}
\Omega_m &=& -{1 \over 3m^2}\, c_2 \, \phi_1^{-2} \, {\rm exp} [ -\phi_1 t], \nonumber \\
\Omega_{\phi} &=& {1 \over 3m^2} \, c_2 \, \phi_1^{-2} \, {\rm exp} [ -\phi_1 t] + 1.
\end{eqnarray}

\noindent Here, it is required $c_2 /c_3 <0$, in order to have a positive values of 
$\Omega_m$. As we have seen, $0< m < 1/2$ $\Longrightarrow$ $m > 0$ and $\phi_1 < 0$, 
which means $\Omega_m$ and $\Omega_{\phi}$ increase exponentially with the time, 
keeping $\Omega_m + \Omega_{\phi}=1$.

\item $\omega \neq -1-m$

\noindent with this condition and $k=0$, the corresponding exact solutions to the set 
of equations (\ref{field11-false-vacuum})-(\ref{field21-false-vacuum}) are

\begin{eqnarray}
\phi(t) &=& \phi_1 \, (c-t)^{\sigma}, \nonumber \\
a(t) &=& a_1 \, (c-t)^{-m \sigma}, \nonumber \\
\lambda(\phi) &=& \lambda_1 \, \phi(t)^{n_1} + {\lambda_2 \over \phi(t)}, \nonumber \\
\rho &=&-{c_1 \over 8\pi}.
\end{eqnarray}

\noindent where 

\begin{eqnarray*}
\sigma &=& {{1-2m} \over {m + \omega + 1} }, \\ 
\phi_1 &=& c_1[c_1^{1/2}(m+\omega+1)]^{\sigma}, \\ 
a_1 &=& \alpha \phi_1^{-m}, \\ 
n_1 &=& {{2 \omega + 3} \over {2m-1}}+1= -{2 \over \sigma}, \\
\lambda_1 &=& -{c_1^{1-n_1} \over 4}{(2m-1)^2 \over {2m + \omega + 1/2}} \Big[ (3+2\omega)
(2m + \omega + 1/2) + (2m-1)(-12m^2 + 3m-6m\omega+3\omega+3/2) \Big], \\ 
\lambda_2 &=& c_2.
\end{eqnarray*}

\noindent This solution corresponds to an extended inflationary model if $m\sigma < 0$. 
In order to have a physical solution, it is required $t<c$. The cosmological constant is 
a decaying function of time: $\lambda \sim t^{-2}+t^{-\sigma}$, with $\sigma >0$. 
But according to the above mentioned condition $m \sigma <0$, it is clear that $m$ 
should be negative.

\noindent This set of solutions is singular just at $t=c$, as we can see from the Ricci 
scalar

\begin{equation}
\Re=6m (2m-1) { {4m^2-3m-\omega-1}\over (m+\omega+1)^2} {1 \over (c-t)^2}.
\end{equation}

\noindent The present deceleration and Hubble parameters for this model, are given as

\begin{equation}
q_0=1-{{m+\omega+1} \over m(2m-1)}, \, \, \, \, \, \, 
H_0=-{m(1-2m) \over {m+\omega+1}}\,(c-t_0)^{-1}, \, \, \, \, \, \, 
t_0=c-{m(1-2m) \over {m+\omega+1}}\, {1 \over H_0}. 
\end{equation}

\noindent According to this results, the expansion of this model takes place in 
accelerated way if $q_0=1+ 1/m\sigma < 0$, but we have seen that $m\sigma < 0$, then 
$\|m \sigma \| < 1$ is the required condition for accelerated expansion. The value of 
$t_0$ clearly depend on the value of $m$ and $\omega$, but we know that, because the 
restriction on $t$, this model is not relevant at present times.

\noindent The corresponding density parameters due to the matter and scalar field for the 
present case, are given by

\begin{eqnarray}
\Omega_m &=& {1 \over 3m^2}{1 \over \sigma^2} \phi_1^{-2/\sigma}\Big[ 8\pi \rho_1-
c_2 \phi(t)^{{4m+2\omega+1} \over {1-2m}} \Big],  \nonumber \\
\Omega_{\phi} &=& {1 \over 3m^2}{1 \over \sigma^2} \phi_1^{-2/\sigma} \Big[ \lambda_1 + 
\lambda_2 \phi(t)^{ {4m + 2\omega + 1} \over {1-2m}} \Big] + {\omega \over 6m^2} + 
{1 \over m}.
\end{eqnarray}

\noindent Both values depend on the free constants, but for all time, it is satisfied that  
$\Omega_m + \Omega_{\phi} = 1$.

\end{enumerate}

\item $k \neq 0$

\noindent In order to solve the set of equations (\ref{field11-false-vacuum})-
(\ref{field21-false-vacuum}) with $k \neq 0$, we shall assume $m=1/2$, then we have 
respectively

\begin{equation}
\label{field12-false-vacuum}
\Big({\phi' \over \phi}\Big)^2(3+2\omega)-{4k \over \alpha^2}=0,
\end{equation}

\begin{equation}
\label{field22-false-vacuum}
{\phi'' \over \phi}(3+2\omega)-{12 k \over \alpha^2}+2{\lambda(\phi) \over \phi}
+2{d\lambda(\phi) \over d\phi}=0.
\end{equation}

\noindent From Eq. (\ref{field12-false-vacuum}) we get

\begin{equation}
\label{field13-false-vacuum}
{\phi' \over \phi}=\Bigg[ {4k \over \alpha^2(3+2 \omega)} \Bigg]^{1/2},
\end{equation}

\noindent solution to this equation is given by

\begin{equation}
\label{sol-phi-false-vacuum}
\phi(\tau)=c_{1} e ^{{2\over \alpha}\sqrt{{k \over {3+2\omega}}}\tau}.
\end{equation}

\noindent From Eq. (\ref{field13-false-vacuum}) we get 

\begin{equation}
\label{field14-false-vacuum}
{\phi'' \over \phi}={4k \over \alpha^2 (3+2\omega)}.
\end{equation}

\noindent We use this last equation in Eq. (\ref{field22-false-vacuum}) from 
which we get a differential equation for $\lambda(\phi)$:

\begin{equation}
\label{de-lambda-false-vacuum}
{d\lambda(\phi) \over d\phi}+{\lambda(\phi) \over \phi}-{4k \over \alpha^2}=0,
\end{equation}

\noindent the solution of this equation reads as follows

\begin{equation}
\label{lambda-false-vacuum}
\lambda(\phi)={2k \over \alpha^2}\phi(\tau)+{c1 \over \phi(\tau)}.
\end{equation}

\noindent From Eqs. (\ref{a-phi}) and (\ref{baro-rho}) with $m=1/2$, we get respectively

\begin{equation}
\label{sol-a-false-vacuum}
a(\tau)=a_{1}e^{-{1\over \alpha}\sqrt{{k\over {3+2\omega}}}\tau},
\end{equation}

\begin{equation}
\label{rho-false-vacuum}
\rho=-{c_{1} \over 8\pi},
\end{equation}

\noindent where $a_{1}\equiv \alpha c_{1}^{-1/2}$. In terms of $t$, this solution 
becomes 

\begin{eqnarray}
\label{sol-false-vacuum-t}
\phi(t)&=& \phi_{1} \, t^{-2}, \nonumber \\
a(t)&=& a_{1} \, t, \nonumber \\
\lambda(\phi)&=& \lambda_{1} \phi(t) +\lambda_{2} \phi(t)^{-1}, \nonumber \\
\rho&=&-{c_{1} \over 8\pi},
\end{eqnarray}

\noindent where $\phi_{1}={\alpha^2 \over k}(3+2\omega)$, $a_{1}=\sqrt{k \over {3+2\omega}}$,
$\lambda_{1}={2k \over \alpha^2}$ and $\lambda_{2}=c_{1}$. This solution has an initial 
singularity, as we can see from the Ricci scalar 
 
\begin{equation}
\label{re-false-vacuum}
\Re=12(2+\omega) \, t^{-2}.
\end{equation}

\noindent $\lambda(\phi)$ increases with the time, in contradiction with the actual observations. 
However, for a particular combination of the today values of the free constants in this model we 
can get a small value of $\lambda_0$

\begin{equation}
\Bigg[ {2 \over \alpha^2}\Big\|{k \over c}\Big\| \phi_1^2 \Bigg]_0^{1/4}
\sim  t_0
\end{equation} 

\noindent The today values of the corresponding deceleration and Hubble parameters are
given by

\begin{equation}
\label{q0-false-vacuum}
q_{0}=0, \, \, \, \, \, \, \, \, H_{0}= {1 \over t_{0}},
\end{equation}

\noindent such that $t_0=1/H_0 \sim 15.05$ Gy.

\end{itemize}

\section{Final remarks}
\label{conclusions}

We have considered a Brans-Dicke scalar-tensor theory, obtaining Friedmann-Robertson-Walker 
cosmological models with time dependent cosmological constant. The time dependence occurs 
in a natural way. Two ansatz were proposed: $a \phi^m = \alpha$, with $\alpha$ constant, and  
$\lambda(t)=\lambda_1 \phi(t)^{n_1}+\lambda_2 \phi(t)^{n_2}$, in order to get exact 
inflationary solutions of the field equations, with a general state equation 
$p=(\gamma - 1)\rho$. Our set of exact solutions depend on the values of $\gamma$, $k$, 
$m$ and $\omega$. 

We classify the exact solutions of each case which we deal, according to the values of the 
free constants of our model. For vacuum with $k=0$ and $m=1/2$, we get a non-relevant 
solution (according to our goal), with $\lambda = 0$. For $k=0$, $m=2/3$, we get a 
singular solution for which the scale factor decreases with the time as $a(t) \sim t^{-1/3}$, 
in accelerated way, but its predicted age is a negative value, then we conclude that this 
model has not physical meaning today. 

Furthermore for vacuum, we get for a flat case, an extended inflationary solution with 
initial singularity, and an exponential inflationary solution without singularity. For a 
not flat case we get a coasting singular solution. In all this models, the values of 
$t_0$ and $\Omega_{\phi}$ (usually called $\Omega_{\Lambda}$) are similar to 
the actual accepted values.

We obtain exact solutions for a general equation of state $p=(\gamma -1 )\rho$. 
In the flat case ($k=0$) we get an extended inflationary solution with initial singularity. 
The expansion of this model occurs in a accelerated way, independently of the equation of state. The values of $t_0$, $\Omega_m$ and 
$\Omega_{\phi}$ depend on the value of the undetermined constants $\gamma$, and $\omega$.

The solution of the non-flat case cannot be expressed in terms of the cosmological time, 
but in terms of the parameter $\tau$. In such  models the initial singularities can be avoided for some values of $\gamma$.

As examples of our general solutions, we calculate the models for a dust and stiff matter 
fluid. For a flat dust model we get a slow extended inflationary solution with acceleration 
and time decaying cosmological constant. This solution has an initial singularity and its 
estimated age is $t_0 \sim 2/H_0$, which is too big according to actual known values. A 
bigger growth rate is required in order to have smaller values of $t_0$ and to be 
consistent with actual observations.

On the other hand, for a non-flat dust model we get a slowly expanding solution 
$a(t) \sim t^2$, with non-constant acceleration or deceleration depending on the relation between our free constants $k/(3+2\omega) >0$ or $k/(3+2\omega)<0$ respectively. The models 
are singular or non-singular depending on the free constants $\alpha$, $k$ and $\omega$.

As another application of our solutions with a general state equation, we consider a 
stiff matter fluid for which $\gamma =2$. The solution of the corresponding flat case is 
a slowly expanding model $a(t) \sim t^{1/2}$, with acceleration, without singularities 
and with a time decaying cosmological ``constant'': $\lambda \sim t^{1/2}$. The 
validity of this model is restricted to some values of $t$. The actual age of the 
Universe predicted by this model, $t_0 \sim 1/2H_0 \sim 7.5$ Gy, with $H_0 \sim 65 \pm 
5$ km ${\rm s}^{-1}{\rm Mpc}^{-1}$; clearly this model is not applicable today.

In the non-flat case for a stiff matter fluid, we get cosmological models for which the 
scale factor grows with the time from a minimum radius, and the cosmological ``constant'' 
decreases with the time, but as in the dust model, we have a set of free  constants, whose  values should determine the characteristics of this model. Thus, 
for $k/(3+2\omega) > 0$ we have a non-constant accelerated model without singularities, 
while for $k/(3+2\omega) < 0$, our solution is a non-constant accelerated model with 
initial singularities and which validity is restricted to some values of $t$. In both 
cases the actual predicted age of the Universe, depend on the numerical value of the 
free constants.

We solve separately the case of a radiation fluid ($\gamma = 4/3$) and a false vacuum 
fluid ($\gamma = 0$). For the radiation fluid we get a coasting model $a(t) \sim t$, 
with a decaying cosmological ``constant'' and with initial singularity, independently 
of the curvature. The value of $t_0 \sim 15.05$ Gy, obtained from this model is in 
fair agreement with actual observations.

For a false vacuum fluid, we obtain a set of solutions which we classify 
depending on the range of values which our free constants may take. For the flat case, we 
get two family of solutions, one of them, corresponds to an exponential inflationary 
model with acceleration and without singularities. Another set of solutions is  
a kind of power law inflationary model $a(t) \sim (c- t)^{\epsilon}$, where $\epsilon$ depend 
on the ``free'' constants $m$ and $\omega$, restricted by physical requirements. For this last 
solution the cosmological constant is a binomial function of $t$, which decreases 
under specific conditions on the free constants.

Additionally, we get a solution for the not flat case of a false vacuum fluid. In such a 
case we get a coasting model $a(t) \sim t$, which has initial singularity and with 
cosmological ``constant''  which is a binomial function of $t$: $\lambda \sim \lambda_1 t^{-2} + 
\lambda_2 t^{2}$. Such a cosmological ``constant'' would increases with the time in contradiction 
with the actual accepted value of $\lambda$. For a particular combination of the today values of 
the free constants in this model, it is possible to obtain small values of $\lambda_0$. The actual 
age predicted by this model is $t_0 \sim 15.05$ Gy, which is again in fair agreement with the 
actual accepted value.

Then, as we can see from the description of our exact solutions, some of them have not 
physical meaning today, some others are restricted to be valid during a specific period. Most 
of  them are valid from an initial singularity until today, predicting an inflationary 
epoch, a cosmological ``constant'' which decreases with the time and the today observed 
acceleration, as well as an actual age of the universe which is in reasonable agreement 
with the actual observations.

\section{Acknowledgments}
Both authors thanks to CONACYT-Mexico by partial financial support.


\begin{thebibliography}{100}
 
\bibitem{weinberg}
S. Weinberg, Rev. Mod. Phys. {\bf 61},  1  (1989).

\bibitem{Ng}
Y. Ng, Int. J. Mod. Phys. {\bf D1},  145  (1992).

\bibitem{bronstein}
M.P. Bronstein, PhysZeit. der Sowjetunion {\bf 3},  73  (1933).

\bibitem{oezer}
M. $\ddot{\rm O}$zer and M. Taha, Mod. Phys. Lett. A {\bf A13},  571  (1998); 
K. Freese, F.Adams, J. Fieman and E. Mottola, Nucl. Phys. B {\bf 287}, 797 (1987); 
P.J.E. Peebles, Astroph. J. {\bf 325}, L17 (1988); 
B. Ratra and P.J.E. Peebles, Phys. Rev. D  {\bf 37}, 3406 (1988); 
M. Reuter and C. Wetterich, Phys. Lett. B {\bf 188} 38 (1987).

\bibitem{chen}
W. Chen and Y.S. Wu, Phys. Rev. D {\bf 41}, 695 (1990).

\bibitem{carvalho}
J.C. Carvalho, J.A.S. Lima and I. Waga, Phys. Rev. D {\bf 46}, 2404 (1992).

\bibitem{capozziello}
S. Capozziello, R. de Ritis and A.A. Marino,  Phys. Lett. {\bf 208A}, 214 (1995).

\bibitem{maty}
J. Matyjasek, Phys. Rev. D {\bf 51}, 4154 (1995).

\bibitem{gasperini}
M. Gasperini, Phys. Lett B {\bf 194}, 347 (1987).

\bibitem{abdel}
A.M. Abdel, Phys. Rev. D {\bf 45}, 3497 (1992).

\bibitem{endo}
M. Endo and T. Fukui, Gen. Rel. Grav. {\bf  8}, 833 (1977).

\bibitem{canuto}
V. Canuto, S.H. Hsieh and P.J. Adams, Phys. Rev. Lett. {\bf 39}, 429 (1977).

\bibitem{kazanas}
D. Kazanas, Astrophs. J. Lett. {\bf 241},  L59 (1980).

\bibitem{bertolami}
O. Bertolami, Il Nuovo Cimento {\bf 93B}, 36 (1986);
M.S. Berman and M.M. Som, Int. J. Theor. Phys. {\bf  29}, 1411 (1990).

\bibitem{lau}
Y.K. Lau, Aust. J. Phys. {\bf 38}, 547 (1985).

\bibitem{lopez}
J.L. Lopez and D.V. Nanopoulos, Mod. Phys. Lett. {\bf  A11}, 1 (1996).

\bibitem{rajeev}
S.G. Rajeev, Phys. Lett. {\bf 125B}, 144 (1983).

\bibitem{hiscock}
W.A. Hiscock, Phys. Lett. {\bf  166B}, 285 (1986).

\bibitem{oezer86}
M. $\ddot{\rm O}$zer and M. Taha, Mod. Phys. Lett. {\bf 171B}, 363 (1986);
R.G. Vishwakarma, Class. Quant. Grav. {\bf 14}, 945 (1997);

\bibitem{calvao}
M.O. Calvao {\it et al}, Phys. Rev. D {\bf 45}, 3869 (1992);
V. M\'endez and D. Pav\'on, Gen. Rel. Grav. {\bf 28}, 679 (1996).

\bibitem{overduin93}
 J.M. Overduin and P.S. Wesson and S. Bowyer, Astrophys. J. {\bf 404}, 1 (1993).

\bibitem{olson}
 T.S. Olson and T.F. Jordan, Phys. Rev. D {\bf 35}, 3258 (1987).

\bibitem{diego}
D. Pav\'on, Phys. Rev. D {\bf 43}, 375 (1991).

\bibitem{maia}
 M.D. Maia and G.S. Silva, Phys. Rev. D {\bf 50}, 7233 (1994).

\bibitem{silveira94}
 V. Silveira and I. Waga, Phys. Rev. D {\bf 50}, 4890 (1994).

\bibitem{torres}
 L.F.B. Torres and I. Waga, Mon. Not. R. Astron. Soc. {\bf 279}, 712 (1996).

\bibitem{silveira97}
 V. Silveira and I. Waga, Phys. Rev. D {\bf 56}, 4625 (1997).

\bibitem{sistero}
 R.F. Sister\'o, Gen. Rel. Grav. {\bf 23}, 1265 (1991).

\bibitem{kalligas}
 D. Kalligas, P. Wesson and C.W.F. Everitt, Gen. Rel. Grav. {\bf 24}, 351 (1992).

\bibitem{arbab}
 A.I. Arbab and A.M. Abdel-Rahman, Phys. Rev. D {\bf 50}, 7725 (1994).

\bibitem{beesham}
 A. Beesham, Phys. Rev. D {\bf 48}, 3539 (1993).

\bibitem{spindel}
 P. Spindel and R. Brout, Phys. Lett. B {\bf 320}, 241 (1994).

\bibitem{waga}
 I. Waga, Astroph. J. {\bf 414}, 436 (1993).

\bibitem{salim}
 J.M. Salim and I. Waga, Class. Quant. Grav. {\bf 10}, 1767 (1993).

\bibitem{lima94}
 J.A.S. Lima and J.C. Carvalho, Gen. Rel. Grav. {\bf 26}, 909 (1994).

\bibitem{wetterich}
 C. Wetterich, Astron. Astrophys. {\bf 30}, 321 (1995).

\bibitem{arbab97}
 A.I. Arbab, Gen. Rel. Grav. {\bf 29}, 61 (1997).

\bibitem{lima-maia}
 J.A.S. Lima and J.M.F. Maia, Phys. Rev. D {\bf 49}, 5597 (1994);
 J.A.S. Lima and M. Trodden, Phys. Rev. D {\bf 53}, 4280 (1996).

\bibitem{kalligas95}
 D. Kalligas, P.S. Wesson and C.W.F. Everitt, Gen Rel. Grav. {\bf 27}, 645 (1995).

\bibitem{moffat}
 J.W. Moffat, Phys. Rev. D {\bf 56}, 6264 (1997).

\bibitem{hoyle}
 F. Hoyle, G. Buerbidge and J.V. Narlikar, Mon. Not. R. Astron. Soc. {\bf 286}, 173 (1997).

\bibitem{john}
 M.V. John and K.B. Joseph, Class. Quant. Grav. {\bf 14}, 1115 (1997).

\bibitem{nesteruk}
E. Gunzig, R. Maartens and A. Nesteruk, Class. Quant. Grav. {\bf 15}, 923 (1998).


\bibitem{overduin}
J. Overduin and F. Cooperstock, Phys. Rev. {\bf D 58},  043506  (1998).

\bibitem{capozziello97}
S. Capozziello and R. de~Ritis, Gen. Rel. Grav. {\bf 29},  1425  (1997).

\bibitem{reasenberg}
R.D. Reasenberg {\it et al}, Astrophys. J. {\bf 234},  L219  (1979).

\bibitem{barrow}
A. Liddle, A. Mazumdar, and J. Barrow, Phys. Rev. D {\bf 58} 027302+ (1998).

\bibitem{dahia}
F. Dahia and C. Romero, gr-qc/9812001; 
M. Susperregi and A. Mazumdar, Phys. Rev. D {\bf 58}, 083512 (1998).

\bibitem{perlmutter98}
S. Perlmutter {\it et.al.}, astro-ph/9812133.

\bibitem{coasting}
L.O. Pimentel and L.M. D{\'\i}az-Rivera, Int. J. Mod. Phys {\bf A}, (to appear).

\bibitem{Pim-Stein}
L.O. Pimentel and J. Stein-Schabes, Phys. Lett. {\bf B 216},  27  (1989). 

\bibitem{guendelman}
E. I. Guendelman, gr-qc 9901017; 
E. I. Guendelman, gr-qc 9901067. 

\bibitem{sethi99}
M. Sethi, A. Batra and D. Lohiya, astro-ph/9903084  .

\bibitem{will}
C. Will, {\em Theory and experiment in gravitational physics} (Cambridge
  University Press, Cambridge, 1981).

\bibitem{johri}
H. Dehnen and O. Obreg\'on, Astrophys. Space Sci. {\bf 17}, 338 (1972);
V. Johri and K. Desikan, Gen. Rel. Grav. {\bf 26},  1217  (1994).

\bibitem{lineweaver}
C. Lineweaver, Astrophys. J. Lett. {\bf 505},  69  (1998).

\bibitem{turner91}
M.S. Turner, Physica Scripta {\bf T36},  167  (1991);
M.S. Turner, {\it In the Critical Dialogues in Cosmology} (World Scientific, Singapore, 1997),
p. 555; L. Krauss and M.S. Turner, Gen. Rel. Grav. {\bf 27} 1137 (1995); 
J.P. Ostriker and P.J. Steinhardt, Nature {\bf 377} 600 (1995).

\bibitem{turner99}
M.S. Turner, in {\it Proceedings of type Ia Supernova: Theory and Cosmology, Chicago 1998} edited by J. Niemeyer and J. Truran (Cambridge University Press, Cambridge U.K. 1999).

\bibitem{chaboyer98}
B. Chaboyer {\em et. al.}, Astrophys. J. {\bf 494},  96  (1998); 
L.M. Macri {\it et. al}, Astrophys. J. (to be published).

\bibitem{carminati}
J. Karminati and R.G. McLenaghan, J. Math. Phys. {\bf 32},  3135  (1991).

\end{thebibliography}
\end{document}